\documentclass[12pt]{article}
\usepackage[dvipdfmx]{graphicx}
\usepackage{amsmath,amssymb}

\title{Finite current stationary states of random 
walks on one-dimensional lattices with aperiodic disorder}

\author{Hiroshi Miki\\
Research Institute for Humanity and Nature,\\
457-4 Motoyama Kamigamo, Kita-ku, Kyoto, 603-8047, Japan}

\begin{document}
\maketitle

\abstract{
Stationary states of random walks with finite induced drift velocity on 
one-dimensional lattices with aperiodic disorder are investigated by scaling 
analysis. Three aperiodic sequences, the Thue-Morse (TM), 
the paperfolding (PF), and the Rudin-Shapiro (RS) sequences, are used to 
construct the aperiodic disorder. These are binary sequences, composed of two 
symbols A and B, and the ratio of the number of As to that of Bs converges to 
unity in the infinite sequence length limit, but their effects on 
diffusional behavior are different. For the TM model, the stationary 
distribution is extended, as in the case without current, and the drift 
velocity is independent of the system size. For the PF model and the RS model, 
as the system size increases, the hierarchical and fractal structure and the 
localized structure, respectively, are broken by a finite current 
and changed to an extended distribution if the system size becomes larger 
than a certain threshold value. Correspondingly, the drift velocity is 
saturated in a large system while in a small system it decreases  
as the system size increases.
}

\newpage

\section{Introduction}
A random walk on a lattice, or more generally on a complex network, is a simple 
stochastic process which describes a classical transport phenomenon in a 
real space or a step-by-step state change in an abstract state space. 
Due to and in spite of the simplicity of the process, various and 
nontrivial properties have been found and analyzed in detail, both 
mathematically and physically\cite{NGvK,Schinazi}.  
 
In the presence of disorder, {\it i.e.}, where the hop probability or rate 
from one site to another is not uniform, the behavior of the random walk is 
strongly modified, especially in lower dimensions, not only quantitatively 
but also qualitatively. When the disorder is random and uncorrelated,
various methods have been developed for analysis. Especially, from methods 
based on renormalization group, many results, some of which are exact, 
have been obtained\cite{ABSO,BG,LDMF,IM}. 
One of the most remarkable results is so-called ultraslow 
diffusion, where the diffusion is strongly suppressed by disorder, and the 
averaged mean-square displacement grows extremely slowly, {\it i.e.} on a 
log-time scale\cite{Sinai,Derrida}. Correspondingly, the stationary state on 
a finite lattice is strongly localized. 

Systems with aperiodic disorder are also interesting  for investigation. 
An aperiodic disorder is generated by a certain set of  
deterministic rules but does not have any periodicity. It is this point that 
distinguishes aperiodic from random uncorrelated disorder. Moreover, 
aperiodic disorder is considered to be intermediate between uniformity or 
periodicity and random disorder. Hence, the study of systems with aperiodic 
disorder is probably a good first step toward understanding systems with more 
general correlated disorder. Generally, it is difficult to construct an 
aperiodic disorder with desirable characteristics. Fortunately, for 
one-dimensional lattice systems various aperiodic disorders can be 
easily constructed with the help of the aperiodic sequences which have been 
investigated mathematically. Note that in addition to the theoretical and 
mathematical interest, systems with aperiodic disorder have been fabricated  
artificially and investigated experimentally\cite{apexp}.
As expected, some results that are unique for 
aperiodically disordered systems have been obtained. One of the most 
remarkable is the appearance of anomalous diffusion, where the mean-square 
displacement grows slowly - less than linearly with time\cite{ITR}. 
Correspondingly, a singular stationary probability distribution with a  
remarkable hierarchical structure appears\cite{Miki}.   
 
If the hop rates do not satisfy a certain condition 
(see Eq.(\ref{zerocurrent})), a finite drift velocity is induced, 
at least in a finite system, and in the stationary state a finite current 
flows through the lattice. As is well known, a finite current forces the 
stationary distribution to be extended. Therefore, it is an interesting 
problem to investigate how the stationary distribution without current will 
be changed by the presence of a finite current - particularly when the 
distribution is localized or singular. In the present paper we use 
scaling analysis to cope with this problem for the cases of lattices with 
aperiodic disorder. As in our previous study\cite{Miki}, 
we consider the aperiodic disorders constructed by the Thue-Morse (TM), 
paperfolding (PF), and Rudin-Shapiro (RS) sequences. 
These three aperiodic sequences have several common properties: (i) They are 
binary sequences, which are composed of two types of symbols, A and B. 
(ii) They are constructed systematically from initial sequences and by 
iteration of specific substitution rules. (iii) The ratio of the number 
of As to that of Bs converges to unity in the infinite sequence length 
limit. Nevertheless, these aperiodic disorders have different effects on 
the diffusional behavior\cite{ITR} and correspondingly on the stationary 
probability distribution\cite{Miki}, since they have difference wandering 
exponents (see Sec. \ref{aperiodicseq}).

We focus on the dependence of the drift velocity and the localization 
structure of the stationary probability distribution on the system size. 
In order to characterize the latter, we use multifractal analysis\cite{HJKPS}, 
as in our previous study\cite{Miki}. 
This approach has been applied to characterize the scaling structure 
of distributions in various systems, including those of the energy dissipation 
in turbulence\cite{CJ,CMJS}, the sidebranch structure of dendrites\cite{MH}, 
and the quantum localization problem\cite{HK}, where the localization property 
of the wavefunction is studied. 

The organization of the rest of this paper is as follows: In Section 2, 
we formulate our model and method for analysis. We describe our 
one-dimensional random walk, give the expressions of the observables, 
and introduce the aperiodic sequences from which the disorder is constructed.
Then we describe the method of the multifractal analysis for the distribution 
on a one-dimensional support, the criterion for localization and the 
finite-size effect. In Section 3, we present our results and a discussion. 
Section 4 is dedicated to our conclusion and future outlook.

\section{Model and method}
\subsection{Random walk on one-dimensional disordered lattice}
Let us consider a random walk on a one-dimensional lattice with only nearest 
neighbor hopping allowed. This process is described by the master equation:
\begin{equation}
\frac{\partial p_j(t)}{\partial t}
= w_{j-1,j}p_{j-1}(t) + w_{j+1,j}p_{j+1}(t)
-(w_{j,j-1}+w_{j,j+1})p_j(t),
\label{mastereq}
\end{equation}
where $p_j(t)$ is the probability for the particle to be on site $j$ at time 
$t$ and $w_{j,k}$ denotes the hop rate for the particle from site $j$ to $k$. 
We impose the periodic boundary condition $p_{j+L} \equiv p_j$ and 
$w_{j+L,k+L}=w_{j,k}$, where $L$ denotes the system size, the number of sites 
on the lattice. Interestingly this master equation is known to be equivalent 
to the transverse-field Ising model\cite{ITR}.

We construct the disorder according to an aperiodic binary sequence, $S$, 
composed of two types of symbols, $A$ and $B$. For example, let us take 
$S=ABAABAABAB\cdots$. For this sequence, the hop rates are assigned as: 
\begin{equation}
w_{j,j+1} = 1, \qquad \text{for all $j$},
\label{forwardrates}
\end{equation}
and
\begin{equation}
w_{j+1,j}=
\begin{cases}
a, &\quad \text{the $j$-th symbol of $S$ is $A$},
\\
b, &\quad \text{the $j$-th symbol of $S$ is $B$}. 
\end{cases}
\label{backwardrates}
\end{equation} 
At least as far as we are concerned with the stationary state, the assignment 
does not lose generality, since the quantities related to the stationary 
state, the probability distribution and the drift velocity, are expressed 
as a function of the ratio $w_{j,j+1}/w_{j+1,j}$\cite{Derrida}.

For the stationary state, $dp_j/dt=0$, the exact expression of the probability 
distribution is obtained in ref. \cite{Derrida} as:
\begin{equation}
p_j = \frac{r_j}{\sum_{k=1}^Lr_k},
\label{probdist}
\end{equation}  
where
\begin{equation}
r_j = 1+\sum_{k=1}^{L-1}\prod_{l=1}^k w_{j+l,j+l-1}.
\label{rj_aux}
\end{equation}   
From this stationary probability, the drift velocity is obtained as 
\begin{equation}
v_d  = \frac{L}{\sum_{j=1}^Lr_j} \left( 1-\prod_{j=1}^L w_{j+1,j} \right).
\label{driftvel}
\end{equation}
If
\begin{equation}
\prod_{j=1}^{L}w_{j+1,j} = 1.
\label{zerocurrent}
\end{equation}
holds, no current exists even in a finite lattice. 
We call this condition the "zero-current condition". For an infinite lattice, 
however, $v_d$ can vanish even if Eq.(\ref{zerocurrent}) does not hold, 
due to the divergence of the denominator of the RHS of 
Eq.(\ref{driftvel}). If the zero-current condition does not hold, 
the finite current tend to force the stationary distribution to be extended 
throughout the lattice. 
Trivially, the stationary distribution is extended if the current is so strong 
that the disorder effect can be considered negligible. 
We are interested in the regime where the effects of current and disorder 
are comparable. 

\subsection{Aperiodic sequences 
\label{aperiodicseq}}
We give a brief review of the binary aperiodic sequences which we consider
for the disorder. Here we describe the properties relevant to the discussion; 
the initial sequence, the substitution rules, and the wandering exponent. 
The readers can refer to the literature, for example, 
refs.\cite{ITR,Miki,Luck,Hermisson} for more details or a general framework 
of aperiodic sequences.

The Thue-Morse (TM) sequence is systematically constructed from the initial 
sequence $S_1=AB$ and the iterative substitution rules $A \rightarrow AB$ and 
$B \rightarrow BA$. This sequence is known to be related to the Koch 
snowflake\cite{MaHol}, which is known to be a fractal curve. 
The length of the sequence of the $n$-th 
generation, $S_n$, is $2^n$. The wandering exponent $\Omega$ is defined as
\begin{equation}
\Delta_n(L) = |N_n(A)-N_n(B)| \sim L^{\Omega},
\label{wanderingexp}
\end{equation}    
where $\Delta_n(L)$ is called the geometric fluctuation of the sequence of 
the $n$-th generation, $L$ denotes the length of the sequence, and $N_n(A)$ 
and $N_n(B)$ denote the numbers of symbols $A$ and $B$, respectively. 
This exponent characterizes how the effect of the disorder changes with the 
system size and plays a crucial role in the diffusional behavior. For the TM 
sequence, $\Omega = -\infty < 0$, and thus the effect of the disorder decreases 
as the system size increases. The diffusional behavior is, even in the case 
without current, essentially similar to that in uniform or periodic systems. 

The paperfolding (PF) sequence is generated by the initial sequence $S_1=AA$ 
and the substitution rules $AA \rightarrow AABA$, $AB \rightarrow AABB$, 
$BA \rightarrow ABBA$, and $BB \rightarrow ABBB$. This sequence is related to 
another fractal, known as the dragon curve\cite{Dekking}. 
The wandering exponent of this sequence is zero. 
In this case, the geometric fluctuation grows 
logarithmically with the system size, $\Delta(L) \sim \log L$, and thus the 
effect of disorder is almost independent of the system size. An anomalous 
diffusion and a multifractal stationary probability distribution are observed 
under the zero-current condition\cite{ITR,Miki}.   

The Rudin-Shapiro (RS) sequence is generated by the initial sequence $S_1=AA$ 
and the substitution rules $AA \rightarrow AAAB$, $AB \rightarrow AABA$, 
$BA \rightarrow BBAB$, and $BB \rightarrow BBBA$. The wondering exponent of 
this sequence is $\Omega=1/2>0$. In this case, the effect of disorder       
grows stronger as the system size increases, and it changes the diffusional 
behavior qualitatively. Without current, a localized stationary 
probability distribution and ultraslow diffusion are observed. Note that the 
value of the exponent $1/2$ coincides with that of the random sequence.

\subsection{Multifractal analysis}
\label{formulation}
To characterize the structure of the stationary probability distribution,  
we use the multifractal analysis method. For a given probability distribution, 
suppose that the support of the distribution is completely covered with 
disjoint patches of size $\epsilon$, and let $p_j(\epsilon)$ be the measure 
assigned to the $j$-th patch. The measure is expected to scale locally as 
\begin{equation}
p_j(\epsilon) \sim \epsilon^{\alpha_j},
\label{localscaling}
\end{equation} 
where $\alpha_j$ denotes the singularity exponent around the $j$-th patch.
Moreover, the number of patches taking the value of the singularity exponent 
between $\alpha$ and $\alpha + d\alpha$, $N(\alpha)$, is expected to scale as 
\begin{equation}  
N(\alpha)d\alpha \sim \epsilon^{-f(\alpha)}d\alpha,
\label{alphascaling}
\end{equation}
where $f(\alpha)$ is, roughly speaking, the fractal dimension of the set of 
patches taking $\alpha$. 

In our study, aperiodic chains are one-dimensional and given through taking 
the limit $L \rightarrow \infty$ for the sequence of finite length. Thus we 
formulate multifractal analysis on a one-dimensional finite lattice and 
evaluate the results of the system of infinite length limit by systematically 
extrapolating the results of the systems of finite size. 

For a given probability measure on a one-dimensional lattice of size $L$, 
$\{P_j\}_{j=1,2,\cdots,L}$, the partition function $Z(q,L)$ is introduced as
\begin{equation}
Z(q,L)=\sum_{j,p_j \ne 0} (p_j)^q.
\label{partition}
\end{equation}
The multifractal exponent for the finite system, $\tau(q,L)$, is defined as
\begin{equation}
\tau(q,L)=-\frac{\log Z(q,L)}{\log L}.
\label{tau_finite}
\end{equation}
The singularity exponent $\alpha=\alpha (q,L)$ and the fractal dimension
$f=f(q,L)$ are obtained through the Legendre transformation:
\begin{eqnarray}
\alpha(q,L) &=& \frac{\partial \tau(q,L)}{\partial q},
\label{alpha_finite}
\\
f(\alpha(q,L)) &=& q\alpha(q,L)-\tau(q,L).
\label{f_finite}
\end{eqnarray} 
Numerically it is better to evaluate them directly, without using numerical 
differentiation. To do this, we follow the method presented in ref.\cite{CJ}, 
summarized below. 

Let us construct a new probability measure $\{\mu_j(q)\}$ from $\{p_j\}$ 
as
\begin{equation}
\mu_j(q) = \frac{(p_j)^q}{\sum_{j=1}^L (p_j)^q}.
\label{mu_construction}
\end{equation}
Then let us define $\zeta(q,L)$ and $\xi(q,L)$ as
\begin{eqnarray}
\zeta(q,L) &=& \sum^L_{j=1} \mu_j(q) \log p_j,
\label{zeta_finite}
\\
\xi(q,L) &=& \sum^L_{j=1} \mu_j(q) \log \mu_j(q),
\label{xi_finite}
\end{eqnarray}
From these we obtain $\alpha(q,L)$ and $f(\alpha(q,L))$ as 
\begin{eqnarray}
\alpha(q,L) &=& -\frac{\zeta(q,L)}{\log L},
\label{alphadef}
\\
f(\alpha(q,L)) &=& -\frac{\xi(q,L)}{\log L}.
\label{fdef}
\end{eqnarray}
It is easy to show, by direct calculation, that the definition 
Eqs.(\ref{alphadef}) and (\ref{fdef}) respectively coincide with 
Eqs.(\ref{alpha_finite}) and (\ref{f_finite}).

Then the evaluation of finite size effect and extrapolation to the infinite 
system size limit are carried out. The value of $\alpha(q)$ is defined 
as the limiting value of $\alpha(q,L)$ as the system size goes to infinity:
\begin{equation}
\alpha(q):=\lim_{L\rightarrow\infty}\alpha(q,L). 
\label{alphalimit}
\end{equation}
From Eqs.(\ref{alphadef}) and (\ref{alphalimit}), 
we expect 
\begin{equation}
\alpha(q)-\alpha(q,L)=\mathcal{O}(1/\log L).
\label{logcorrection}
\end{equation}
From this we can evaluate the value of $\alpha(q)$ from the plot of 
$\alpha(q,L)$ against $1/\log L$ and the extrapolation to 
$1/\log L \rightarrow 0$. 
The fractal dimension $f(q)$ is evaluated similarly.

Let $\alpha_{\rm min}$ and $f_{\rm min}$ be $\alpha(q \rightarrow \infty)$ and 
$f(q \rightarrow \infty)$, respectively. 
We can read the localization property of a given distribution from 
these exponents\cite{HK}. 
For an extended distribution, the multifractal spectrum of finite  
systems, which draws a curve in the $\alpha-f$ plane, converges to a single 
point, $\alpha=f=1$ in the limit as $L\rightarrow\infty$. For a localized 
distribution, both $\alpha_{\rm min}$ and $f_{\rm min}$ converges to 0. 
For a singular distribution, $\alpha_{\rm min}$  converges to a certain finite 
value and $f_{\rm min}$ may converge to 0 or a finite value, depending on the 
case. In our discussion below, we investigate the scaling properties of only 
$\alpha_{\rm min}$, since that is sufficient for determining the 
localization property. 

\section{Results and discussion}
We are interested in the cases where the local preferential hopping 
direction depends on the disorder. Thus we restrict ourselves to the 
cases with $a<1$, $b>1$, and $ab<1$, unless otherwise noted.
The condition $ab<1$ means that the current is in the positive direction 
(at least in the $L \rightarrow \infty$ limit). We expect that the 
characteristic properties are invariant under $a \leftrightarrow b$, 
since for the underlying aperiodic sequences, the ratio 
of the number of As to that of Bs converges to unity in the infinite  
length limit. We can quantify the degree of the breaking of the 
zero-current condition as $1-ab$. 

\subsection{Thue-Morse(TM) model}
The stationary probability distribution of the TM model is extended even 
when there is no current\cite{Miki}, since the wandering exponent of this 
sequence is negative. Therefore, in the case with finite current, 
the extendedness of the distribution is still maintained, and only a 
quantitative difference is observed.
\begin{figure}
\begin{center}
\includegraphics[width=12cm]{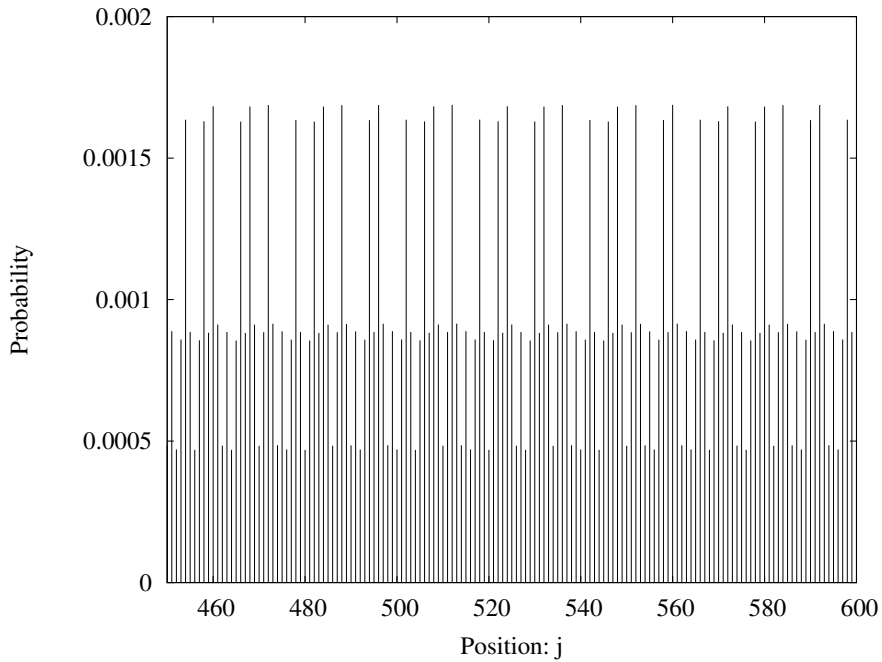}
\caption{Stationary probability distribution $\{p_j\}$ for the TM model 
with $a=0.5$, $b=1.8$, and $L=1024$. Only the region $450 \ge j \ge 600$ 
is shown for visibility. The distribution is extended.  
\label{tmdist}}
\end{center}
\end{figure}
Figure \ref{tmdist} shows the stationary distribution of the TM model with 
$a=0.5$, $b=1.8$, and $L=1024$. The measures are fluctuating around one of 
the three values, $1/C$, $a/C$, and $1/aC$, 
where $C=2^{n-1}+(a+a^{-1})2^{n-2}$ and $n=\log_2L$, which are the measures in 
the case without current. 
 
\begin{figure}
\begin{center}
\includegraphics[width=12cm]{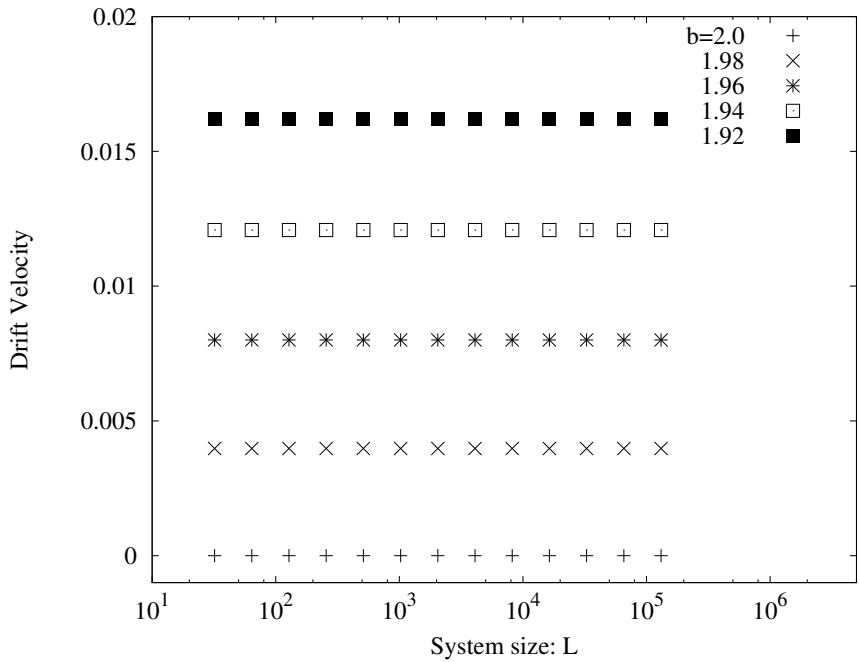}
\caption{Plots of the drift velocity $v_d$  for the TM model for several 
values of $b$ with $a$ fixed at 0.5.
\label{tmvelplot}}
\end{center}
\end{figure} 
The drift velocity is numerically almost independent of the system size.
(See Fig.\ref{tmvelplot}.) 
This is the consequence of the negativeness of the wandering exponent and 
the fact that for the TM sequence, the numbers of A and B exactly coincident 
in any generation $n$. We also note that the drift velocity is proportional 
to the degree of the breaking of the zero-current condition:
\begin{equation}
v_d \propto 1-ab,
\label{tmvel}
\end{equation}  
as in cases where the arrangement of A and B is periodic.

\subsection{Paperfolding(PF) model}
\begin{figure}
\begin{minipage}{0.8\hsize}
\begin{center}
\includegraphics[width=12cm]{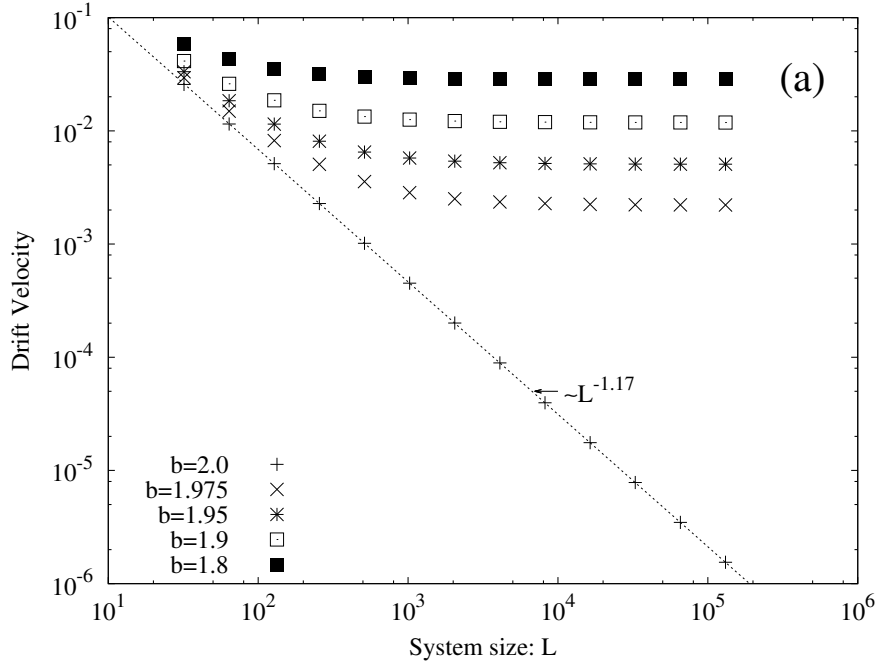}
\end{center}
\end{minipage}
\begin{minipage}{0.8\hsize}
\begin{center}
\includegraphics[width=12cm]{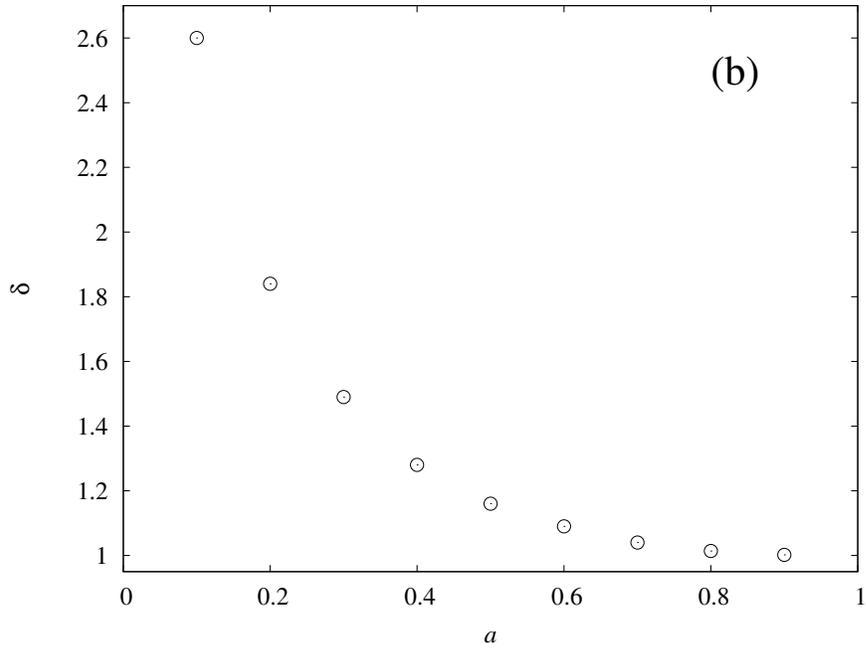}
\end{center}
\end{minipage}
\caption{(a) Plots of the drift velocity $v_d$ as a function of the system 
size $L$ for the PF model for several values of $b$ 
with $a$ fixed at 0.5. (b) Dependence 
of the exponent $\delta$ (see Eq.(\ref{veldecaypf})) on $a$. It converges 
to unity in the limit $a \rightarrow 1$. 
\label{velplotpf}}
\end{figure}

Before analyzing the structure of the stationary distribution in detail, 
we consider the dependence of the drift velocity $v_d$ on the system size.
Figure \ref{velplotpf}(a) shows this plot for several values of $b$ 
with $a=0.5$ fixed.
First, for $b=2.0$, where the zero-current condition is satisfied in the 
limit of infinite system size, it is observed that the drift velocity decays 
algebraically:
\begin{equation}
v_d \sim L^{-\delta},
\label{veldecaypf}
\end{equation}
where the exponent $\delta$ depends on $a$. Figure \ref{velplotpf} (b) shows 
the dependence of $\delta$ on $a$. It converges to unity in the limit as 
$a \rightarrow 1$, the homogeneous limit.

If the zero-current condition does not hold even in the limit of infinite 
system size, the drift velocity decays for a small system , and it is 
saturated and converges to a certain finite value for a large system. 
\begin{figure}
\begin{minipage}{0.8\hsize}
\begin{center}
\includegraphics[width=12cm]{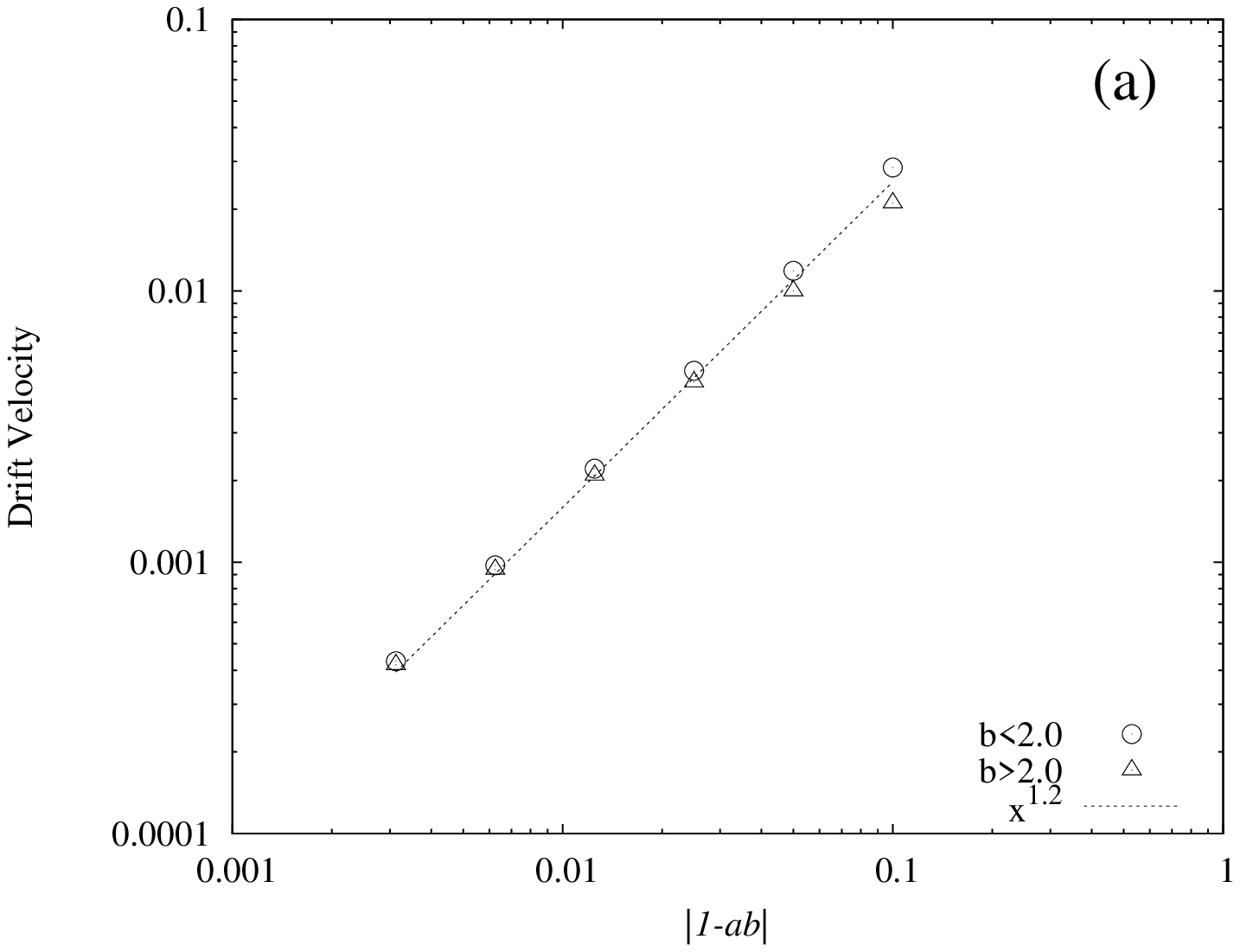}
\end{center}
\end{minipage}
\begin{minipage}{0.8\hsize}
\begin{center}
\includegraphics[width=12cm]{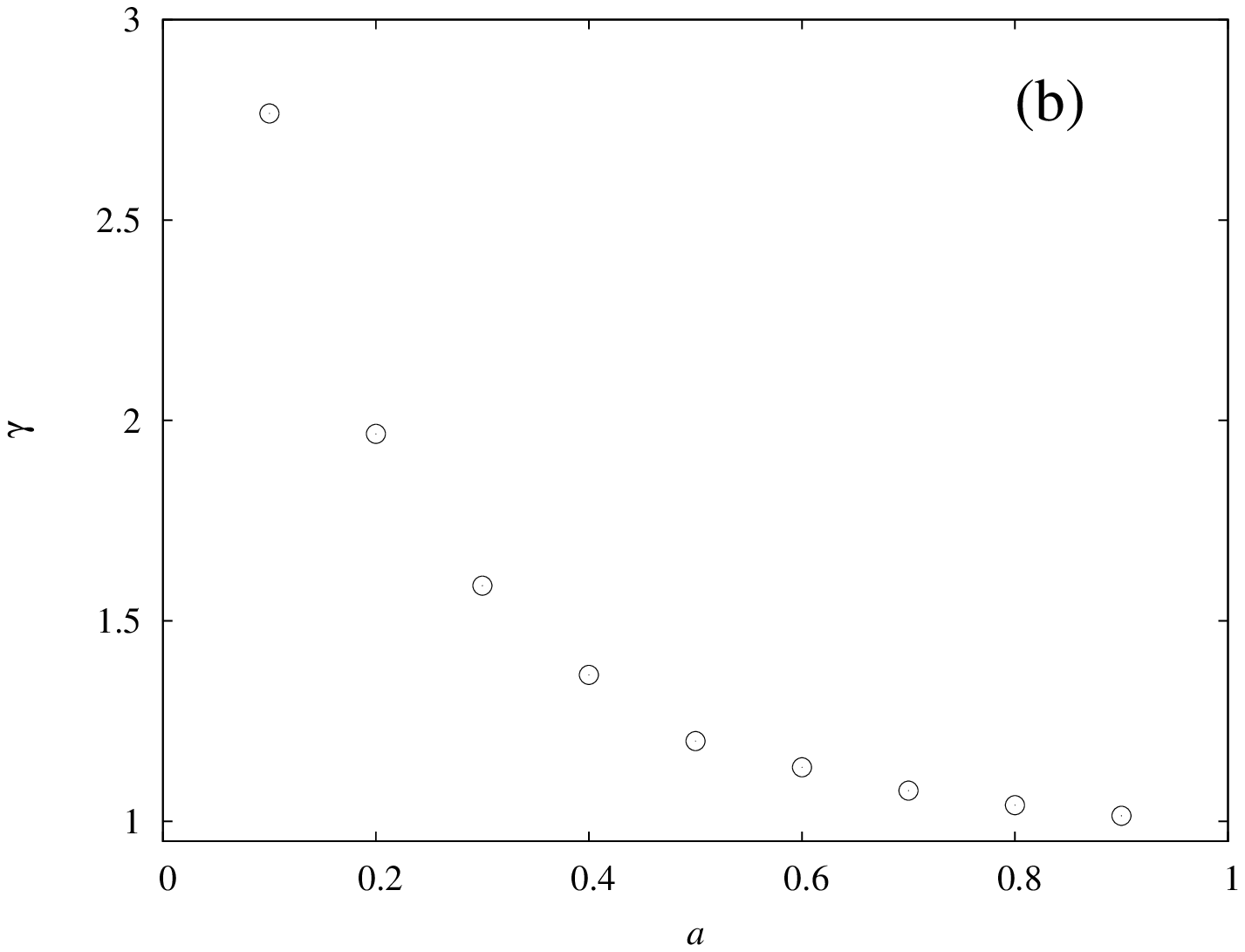}
\end{center}
\end{minipage}
\caption{(a)Plot of the converged drift velocity for the PF model 
in the limit as 
$L \rightarrow \infty$ against $|1-ab|$, the degree of the breaking of the 
zero-current condition. (b)Plot of the exponent in Eq.(\ref{velbrokenpf}) 
as a function of $a$. It converges to unity in the limit $a \rightarrow 1$.   
\label{velbpf}}
\end{figure}
Figure \ref{velbpf}(a) shows the dependence of the drift velocity on the 
degree of the breaking of the zero-current condition, $|1-ab|$ with $a=0.5$
fixed. It is observed to be a power-law
\begin{equation}
|v_d| \sim |1-ab|^{\gamma}.
\label{velbrokenpf}
\end{equation}
regardless of the sign of $1-ab$. 
Figure \ref{velbpf}(b) shows the dependence of the exponent $\gamma$ 
on $a$. 

For the PF model without current, the stationary probability distribution is 
called ''singular'', neither extended nor localized. It has a hierarchical 
structure and is characterized by multifractal spectrum\cite{Miki}. 
\begin{figure}
\begin{center}
\includegraphics[width=12cm]{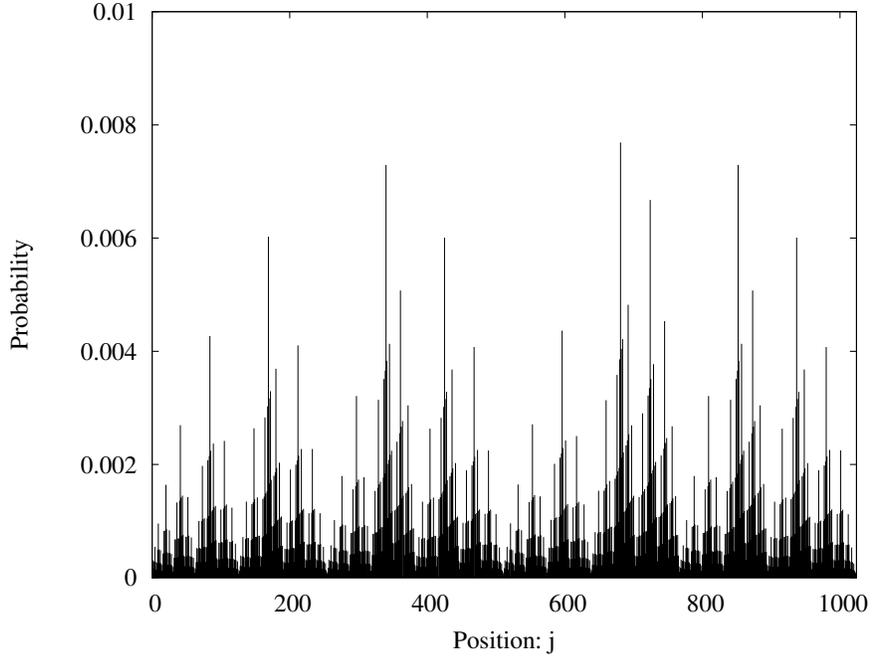}
\caption{Stationary probability distribution for the PF model with $a=0.5$, 
$b=1.9$, and $L=1024$. The hierarchical structure of the distribution in the 
case without current is being broken and the distribution is being extended. 
\label{pfdist}}
\end{center}
\end{figure}
Figure \ref{pfdist} shows the stationary distribution of the PF model with 
$a=0.5$, $b=1.9$ and $L=1024$. It is observed that the hierarchical structure 
is gradually being broken and the distribution is forced to be extended.

\begin{figure}
\begin{center}
\includegraphics[width=12cm]{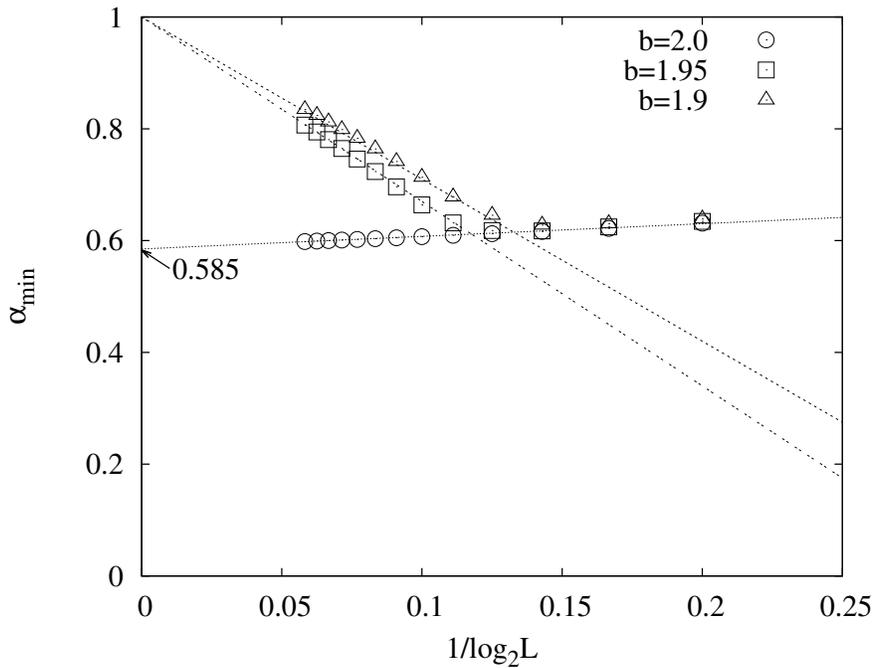}
\caption{Plots of $\alpha_{\rm min}(L)$ against $1/n=1/\log_2L$ for the 
stationary distribution of the PF model with $a=0.5$ and several values of 
$b$. A shift is observed from the scaling behavior of a singular 
distribution to that of an extended distribution.  
\label{aminplot_pf}}
\end{center}
\end{figure}
Figure \ref{aminplot_pf} shows the plots of $\alpha_{\rm min}(L)$ for several 
values of $b$ with $a=0.5$ fixed against $1/n=1/\log_2L$ for the stationary 
distribution. For $b=2.0$, the zero-current condition is satisfied in the 
infinite system size limit. The plot is linear and converges to a finite 
value, $\alpha_{\rm min}=0.585$, in the limit as $\/\log_2L \rightarrow 0$ 
({\it i.e.} $L \rightarrow \infty$). This means that the distribution is 
singular. Meanwhile, for $b=1.95$ and $1.9$, where the zero-current condition 
does not hold, a different behavior is observed. For a small system, 
the plots are on a line corresponding to a singular distribution, as in the 
case with $b=2.0$. However, for a large system, the plot shifts  
to a line corresponding to an extended distribution, which converges to unity 
as $1\/\log_2L \rightarrow 0$. The shift takes place at a smaller system 
size for $b=1.9$ than for $b=1.95$, since the degree of the breaking of the 
zero-current condition of the former is larger than that of the latter. 
In both cases, in the infinite system size limit, the stationary distribution 
is extended.  

Note that the system size at which the shift takes place roughly 
corresponds to the size at which the drift velocity is saturated. 
For example, for $b=1.95$, the shift and saturation take place at 
$L \sim 10^3$ ($n=9-10$). This length is considered as a correlation 
length, within which the disorder is effective. 

From these results, we can conclude that any small breaking of the 
zero-current condition induces nonzero drift velocity and forces the 
stationary distribution to be extended.

\subsection{Rudin-Shapiro(RS) model}
First we consider the case where the zero-current condition is satisfied 
in the infinite system size limit.
\begin{figure}
\begin{center}
\includegraphics[width=12cm]{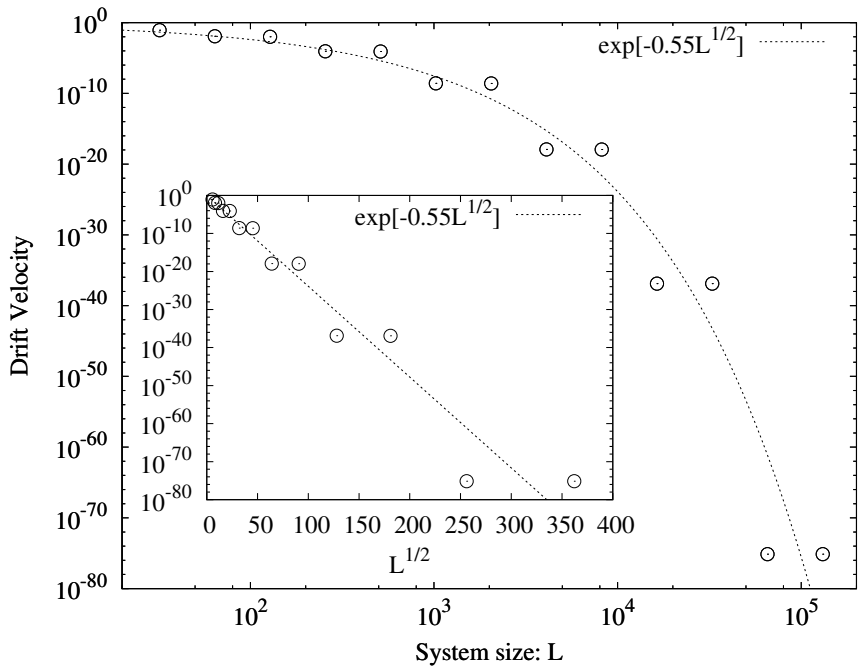}
\caption{Dependence of the drift velocity of the RS model on the system 
size with $a=0.5$ and $b=2.0$. (Inset) The same plot with $\sqrt{L}$ 
as the abscissa. 
\label{veldif_rs_unbiased}}
\end{center}
\end{figure}
Figure \ref{veldif_rs_unbiased} shows the dependence of the drift velocity 
$v_d$ on the system size $L$ in the case with $a=0.5$ and $b=2.0$. It decays 
very rapidly as the system size increases and it is fitted well by 
\begin{equation}
v_d(L) \sim \exp[-cL^{1/2}],
\label{rsveldecay}
\end{equation}
where $c \sim 0.55$.

\begin{figure}
\begin{center}
\includegraphics[width=12cm]{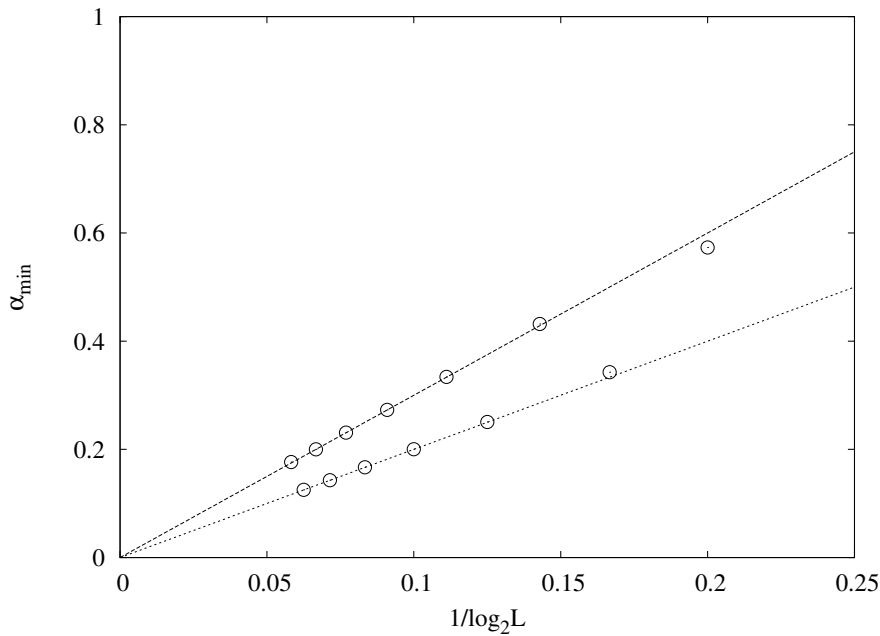}
\caption{Plot of $\alpha_{\rm min}(L)$ against $1/n=1/\log_2L$ for the 
stationary distribution for the RS model with $a=0.5$ and $b=2.0$. 
The upper line corresponds to the group with odd $n$ and the lower 
to the group with even $n$.  
\label{amin_rs_unbiased}}
\end{center}
\end{figure}
Figure \ref{amin_rs_unbiased} shows the dependence of $\alpha_{\rm min}$ on 
$1/n=1/\log_2L$ for the stationary distribution in the same case. There is 
a parity dependence found, {\it i.e.} the plots are classified into two 
groups - of even $n$ and odd $n$. Anyway, although the plots in different 
groups are on different lines, both converges to zero in the infinite system 
size limit, $1/\log_2L \rightarrow 0$. This fact indicates that the 
distribution is localized. This scaling behavior is a little different from 
that of the case where the zero-current condition is always satisfied in  
finite size lattice. In that case, the line corresponding to the group of 
odd $n$ is merged into that to the group of even $n$ for sufficiently large 
system size\cite{Miki}.   

\begin{figure}
\begin{center}
\includegraphics[width=12cm]{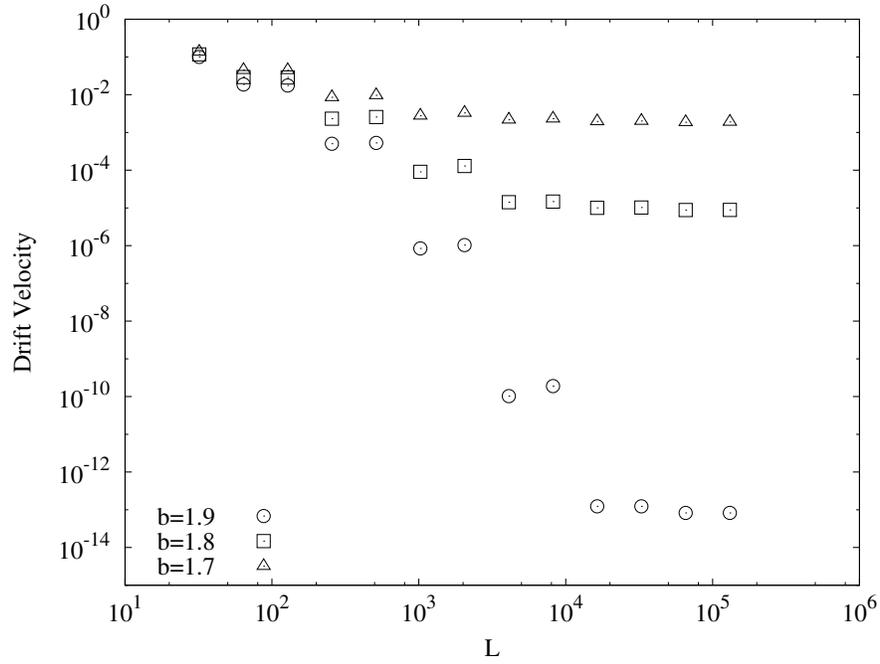}
\caption{ Plots of the drift velocity $v_d$ against the system size $L$ 
for the RS model with $a=0.5$ fixed and several values of $b$.
\label{veldif_rs_biased}}
\end{center}
\end{figure}
Figure \ref{veldif_rs_biased} shows the dependence of the drift velocity on 
the system size for several values of $b$ with $a=0.5$ fixed, where the 
zero-current condition does not hold. In these cases, if the system is 
small, as the system size increases, the drift velocity decreases. 
After that the system becomes larger than a certain threshold value, 
the drift velocity is saturated and converges to a certain finite value. 
For fixed $a$, as the degree of the breaking of the zero-current condition, 
$|1-ab|$, grows larger, the threshold value for the system size becomes 
smaller and the drift velocity converges to a larger value. 
However, unfortunately, it is quite difficult to obtain the value to which 
the drift velocity converges as a function of the degree of the breaking of 
the zero-current condition, since the smaller the degree of the breaking, 
the faster increases the threshold system size at which the drift velocity 
is saturated and converges, and the smaller the value to which the velocity 
converges.      

\begin{figure}
\begin{center}
\includegraphics[width=12cm]{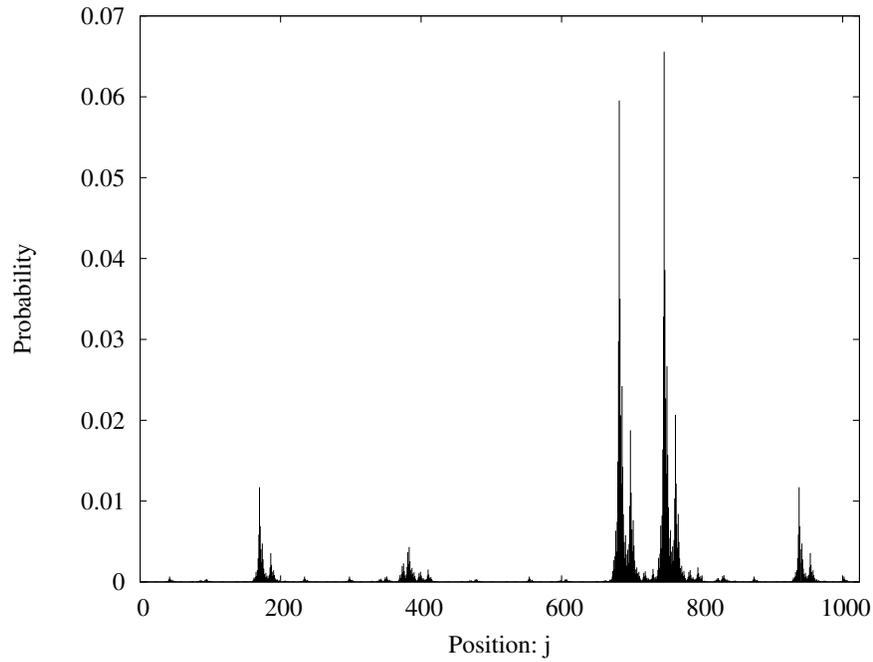}
\caption{Stationary probability distribution for the RS model with $a=0.5$, 
$b=1.7$, and $L=1024$. Several high density regions can be observed. 
\label{rsdist}}
\end{center}
\end{figure}
Figure \ref{rsdist} shows the stationary probability distribution for 
$a=0.5$, $b=1.7$, and $L=1024$. Several high-density regions are observed 
and therefore the distribution is forced to be extended by a finite current.  

\begin{figure}
\begin{center}
\includegraphics[width=12cm]{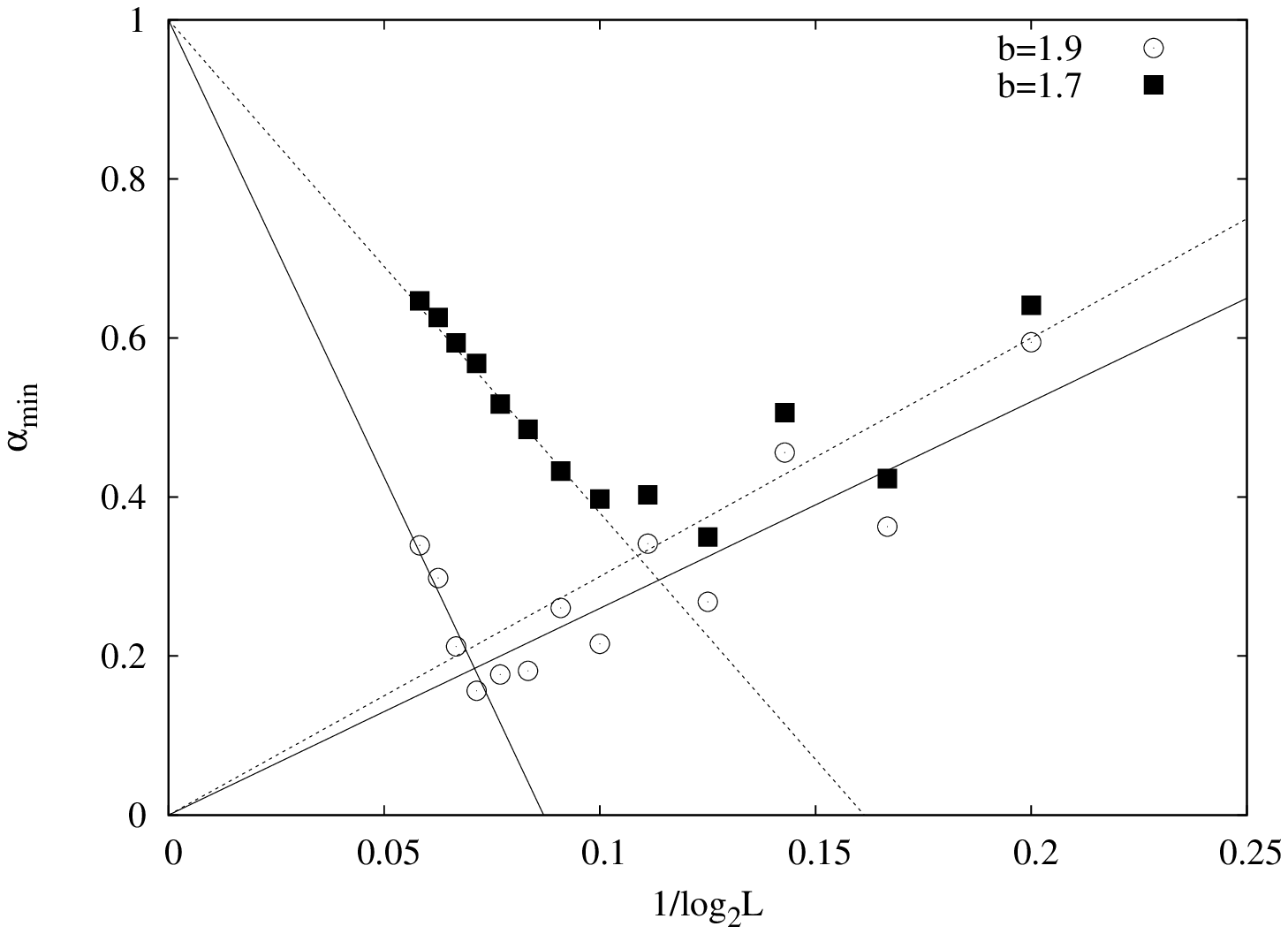}
\caption{Plots of $\alpha_{\rm min}(L)$ against $1/n=1/\log_2L$ for the 
stationary distribution of the RS model with $a=0.5$ and $b=1.9$ and $1.7$. 
A shift is observed from the scaling behavior of a localized distribution 
to that of an extended distribution.   
\label{amin_rs_biased}}
\end{center}
\end{figure}
Figure \ref{amin_rs_biased} shows the dependence of $\alpha_{\rm min}$ on
$1/n=1/\log_2L$ for $a=0.5$ and $b=1.9$ and $1.7$. For a small system, 
$\alpha_{\rm min}$ approaches zero as $1/n \rightarrow 0$. 
This is the characteristic 
scaling behavior of a localized distribution, as in the case where the 
zero-current condition is satisfied in the infinite system size limit. 
Then, the plot shifts towards unity after the system becomes larger than 
a certain threshold value. This corresponds to the scaling behavior of an 
extended distribution. Note that, as seen in the PF model, the system size 
at which the shift of the scaling behavior of $\alpha_{\rm min}$ takes 
place is roughly the same as that at which the saturation of the drift 
velocity begins, and the larger the degree of the breaking of the 
zero-current condition becomes, the smaller the threshold for the system 
size is.

It is known that, for a system with random binary disorder, the drift 
velocity may vanish in the infinite system size limit, 
even if the zero-current condition is not 
satisfied\cite{BG, Sinai, Derrida}. The drift velocity vanishes if 
(1) the ratio of the two symbols does not converge to unity and (2) there 
is a sufficiently large weight corresponding to the symbol of fewer 
occurences.   
Since the wandering exponent of the RS sequence coincides with that of a 
random binary sequence, we conclude that in the RS model a finite 
drift velocity is induced by any small breaking of the zero-current condition 
and the drift velocity breaks the localized distribution when the system is 
large.        

\section{Conclusion and outlook}
We have investigated the stationary states of random walks on one-dimensional 
lattices with aperiodic disorder, where a finite current flows through the 
system. Binary aperiodic sequences were used to construct the disorder, where 
the ratio of the number of As to that of Bs converges to unity in the limit of 
infinite sequence length. We concluded that no matter how little the 
zero-current condition is broken in the limit of infinite system size, 
a finite drift velocity is induced and for a large system the stationary 
distribution is extended, irrespective of the wandering exponent of the 
underlying aperiodic sequence, which affects the diffusional behavior. 
This conclusion may sound natural, or even trivial, since irrespective of 
the property of the diffusion - normal, anomalous or ultraslow - it is 
well known that diffusion is dominant in small systems and drift is dominant 
in large systems. However, for randomly disordered cases, it has been known 
that under certain conditions, the drift velocity vanishes, even though 
the zero-current condition is broken in the infinite system size limit, 
if the ratio of the number of As to that of Bs does not converge to 
unity\cite{BG,Derrida}. In this sense, The symmetry between A and B is 
a special condition. 
It has not yet been clear how the drift velocity and the stationary 
distribution behave in cases with asymmetrical aperiodic disorder. 
We expect that the differences between random disorder and aperiodic 
disorder may be much clearer in those cases. This is a problems to be 
investigated in the future.  
\\ 
\\ 
{\Large \bf Acknowledgments}

This research was supported by the initiative-based project E-05 
"Creation and Sustainable Governance of New Commons through Foundation of 
Integrated Local Environmental Knowledge (ILEK)", Research Institute for 
Hunanity and Nature (RIHN).

\end{document}